\documentclass[10pt,conference]{IEEEtran}
\usepackage{graphicx} 
\usepackage[utf8]{inputenc}
\usepackage{cite}
\usepackage{algorithm}
\usepackage{amsmath, amssymb}
\usepackage{subfiles}
\usepackage{url}
\usepackage{hyperref}
\usepackage{booktabs}
\usepackage{graphicx}
\usepackage{caption}
\usepackage{subcaption}
\usepackage{textcomp}
\usepackage{bm}
\usepackage{xcolor}
\usepackage{amsthm}
\usepackage{algpseudocodex}
\usepackage{arydshln}
\usepackage{pifont}

\usepackage[draft, commentmarkup=todo,
  todonotes={textsize=tiny, textwidth=0.83in}]{changes}

\definechangesauthor[name={Ilya Safro}, color=yellow]{IS}

\newif\ifisjournal
\isjournaltrue

\title{QAdaPrune: Adaptive Parameter Pruning For Training Variational Quantum Circuits}
\author{\IEEEauthorblockN{Ankit Kulshrestha}
\IEEEauthorblockA{\textit{Fujitsu Research of America} \\
Santa Clara, CA\\
akulshrestha@fujitsu.com}
\and
\IEEEauthorblockN{Xiaoyuan Liu}
\IEEEauthorblockA{
\textit{Fujitsu Research of America}\\
Santa Clara, CA \\
xliu@fujitsu.com}
\and
\IEEEauthorblockN{Hayato Ushijima-Mwesigwa}
\IEEEauthorblockA{
\textit{Fujitsu Research of America}\\
Santa Clara, CA \\
hayato@fujitsu.com}

\and
\IEEEauthorblockN{Bao Bach}
\IEEEauthorblockA{\textit{Department of Computer and Information Sciences}\\
\textit{University of Delaware}\\
Newark, DE\\
baobach@udel.edu}
\and
\IEEEauthorblockN{Ilya Safro}
\IEEEauthorblockA{\textit{Department of Computer and Information Sciences} \\
\textit{University of Delaware}\\
Newark, DE \\
isafro@udel.edu}}

\date{April 2023}

\DeclareMathOperator*{\argmin}{arg\,min}

\newcommand{\ket}[1]{\ensuremath{|#1 \rangle}}
\newcommand{\bra}[1]{\ensuremath{\langle #1 |}}

\begin{document}

\maketitle

\begin{abstract}
In the present noisy intermediate scale quantum computing era, there is a critical need to devise methods for the efficient implementation of gate-based variational quantum circuits. This ensures that a range of proposed applications can be deployed on real quantum hardware. 
The efficiency of quantum circuit is desired both in the number of trainable gates and the depth of the overall circuit. The major concern of barren plateaus has made this need for efficiency even more acute. The problem of efficient quantum circuit realization has been extensively studied in the literature to reduce gate complexity and circuit depth. Another important approach is to design a method to reduce the \emph{parameter complexity} in a variational quantum circuit. Existing methods include hyperparameter-based parameter pruning which introduces an additional challenge of finding the best hyperparameters for different applications. In this paper, we present \emph{QAdaPrune} - an adaptive parameter pruning algorithm that automatically determines the threshold and then intelligently prunes the redundant and non-performing parameters. We show that the resulting sparse parameter sets yield quantum circuits that perform comparably to the unpruned quantum circuits and in some cases may enhance trainability of the circuits even if the original quantum circuit gets stuck in a barren plateau.\\ \noindent{\bf Reproducibility}: The source code and data are available at \url{https://github.com/aicaffeinelife/QAdaPrune.git}
\end{abstract}
\section{Introduction}\label{sec:intro}

\begin{figure*}
\centering
\includegraphics[width=1.65\columnwidth]{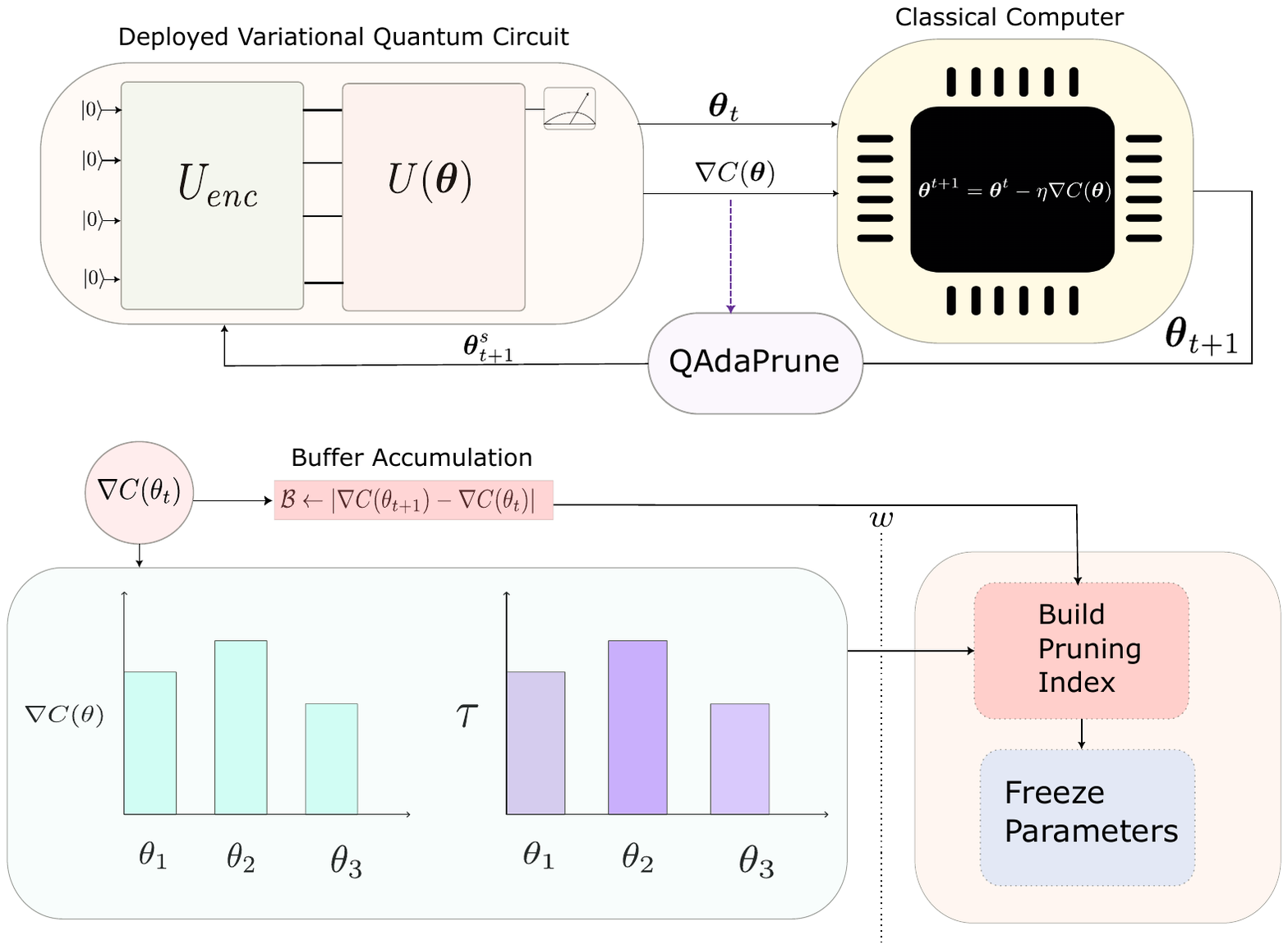}

\caption{Overview of a deployed VQC (top) and the QAdaPrune algorithm(bottom). The algorithm accepts a vector of parameters $\bm{\theta}_t$ and applies an adaptive pruning procedure to return pruned parameters. The procedure introduces minimal overhead in the training process of a VQC.}
\label{fig:qadaprune}
\end{figure*}

Quantum computing is an emerging discipline that aims to leverage quantum mechanical effects to process information faster than currently possible with a conventional computer. While a fully functioning quantum computer is yet to be realized, there exists a generation of quantum computers known as noisy intermediate scale quantum (NISQ) computers that simulate the properties of a quantum computer and allow development of algorithms that can obtain more speedup when a fully functioning quantum computer is finally built. One of the dominant paradigms in current setting is that of Variational Quantum Algorithms (VQAs)~\cite{cerezo2019variational} that use a hybrid computing setup to obtain solutions to various optimization problems. Some examples of their applications include approximating solutions to such NP-hard problems as graph maxcut~\cite{farhi2014qaoa_for_bounded}, minimum graph partitioning \cite{ushijima2021multilevel}, community detection \cite{shaydulin2019network,liu2022layer}, estimating the ground state energy of molecules~\cite{VQE}, portfolio optimization \cite{herman2023quantum}, pattern recognition~\cite{farhi2018classification}, unsupervised data compression~\cite{romero2017quantum} and even circuit architecture search~\cite{du2022quantum}.

In this paper, we consider using a class of variational quantum circuits (VQC)s called Quantum Neural Networks (QNNs)~\cite{farhi2018classification} for pattern classification tasks. In theory, QNNs can process large datasets faster and scale up  beyond current limits of Deep Neural Networks (DNNs) but  issues in quantum computers (discussed later in this paper) prevent them for being useful for high dimensional and large sample size datasets. Our work aims to bridge the gap between the purported theoretical benefits of QNNs and their practical applicability by devising a novel parameter pruning procedure that allows for deployment of deeper QNNs on existing quantum hardware. 

In the QNN setup, a vector of parameters $\bm{\theta}$ indicates the rotation angles of quantum ``gates" that manipulate information contained in an input quantum state $\ket{\psi}_{in}$. For pattern classification tasks, the output of the deployed quantum circuit is a probability vector that indicates the confidence of the QNN over different classes. These circuits are trained using a hybrid regime where the loss and the gradient are computed on a quantum device and the parameters updated on a classical computer using gradient descent(Figure~\ref{fig:qadaprune}). The only way to compute gradient of loss w.r.t. parameters is to use the parameter-shift rule~\cite{schuld2019evaluating}. The most significant issue in the deployment of quantum circuits is the barren plateau effect~\cite{mcclean2018barren} due to  which the expected value of gradient converges to zero and its variance decays exponentially with the number of qubits. \emph{This is a critical obstacle in deploying many AI, machine learning or optimization algorithms on quantum systems that prevents them to be applied on practically interesting data (even if the data is not large).} 
Some methods have been proposed to work around the short depth limit of quantum circuits. These include transpiling of a complex simulator circuit into an efficient gate level physical circuit~\cite{rakyta2022efficient}, generating efficient quantum circuits from a known ``supercircuit"~\cite{du2022quantum} and pruning parameters to reduce the effective gate depth of the circuit~\cite{wang2022qoc, hu2022quantum, sim2021adaptive}. Out of these methods, parameter pruning is the most versatile and understudied method for improving the performance to resource ratio of quantum circuits. Their versatility stems from the fact that they can be used standalone or as pre-processing methods for a downstream transpiler. In the latter case, they can help the transpiler understand and eliminate redundant circuit paths and gates to obtain efficient and performant circuits. Existing pruning algorithms suffer from two major shortcomings. First, they rely on a global threshold across all parameter components. Second, they rely on heuristics to prune/regrow parameters. These heuristics often require tunable hyperparameters and are thus implicitly reliant on finding the optimal set of hyperparameters for the given data and circuit. The overhead introduced by the necessity of hyperparameter tuning thus makes these pruning algorithms undesirable for applying to large depth circuits. 

Motivated by the importance of parameter pruning and the shortcomings of existing methods we make the following contributions:

\begin{itemize}
    \item We propose the \texttt{QAdaPrune} algorithm that adaptively learns the pruning threshold for each parameter component and optimizes pruning procedure to produce accurate and resource-efficient quantum circuits.

    \item We demonstrate that using a better pruning criteria (i.e. approximated Hessian of cost function) can yield a better parameter pruning.

    \item We demonstrate the performance of our algorithm in pruning quantum neural networks on two different classification datasets. 
\end{itemize}

%
%
%
%
\section{QAdaPrune Algorithm}\label{sec:theory}
\begin{algorithm}
\caption{QAdaPrune Algorithm
    \label{alg:qadaprune_alg}}
\begin{algorithmic}[1]
\Require $w$: profiling window size, $k$: parameters to retain, $T$: total number of gradient descent steps, $\eta$: learning rate, $\bm{\theta} \in \mathbb{R}^{N}$: initial paramters.

\Ensure $\bm{\tau} \gets \frac{1}{n}$, $\bm{s} \gets \{ \}$

\State $\mathcal{B} \gets \nabla C(\bm{\theta})$ \Comment{Initial state of buffer}
\State $t \gets 0$
\While{$t \leq T$}
    \State $\bm{\tau}' \gets \bm{\tau} ( 1 - |\nabla C(\bm{\theta}_t)|)$ \Comment{Threshold adjustment}
    \State $\bm{\theta}_{t+1} \gets \bm{\theta}_t - \eta \nabla C(\bm{\theta}_t)$ 
    \State $\mathcal{B} \gets \mathcal{B} + |\nabla C(\bm{\theta}_t) - \nabla C(\bm{\theta}_{t+1})|$ \Comment{Gradient accumulation in buffer}
    \If{$t\bmod \kappa.w = 0$} 
        \State $\bm{s} \gets$ \Call{SALIENCY}{$\bm{\tau}', \mathcal{B}$}
        \State $\mathcal{B} \gets \nabla C(\bm{\theta}_t)$ \Comment{Reset to current gradient}
    \EndIf
    \State $\bm{\theta}^s_t \gets$ \Call{PRUNE\_PARAMETERS}{$\bm{\theta}_t, \mathcal{B}, \bm{s}, k$}
\EndWhile
\State \Return $\bm{\theta}^s_*$ \Comment{The optimized and pruned parameters.}

    
    
        
\end{algorithmic}
\end{algorithm}

\ifisjournal
In order to ease the exposition and explain the key components of the proposed algorithm, we first begin by reviewing the VQA framework for a multi-layer VQC deployed on a quantum device. We assume that we have access to an initial state $\ket{0}$ and a VQC defined as
\begin{equation}
    U(\bm{\theta}) = \prod_{l=1}^{L} U(\theta^{l}), 
    \label{eq:vqc}
\end{equation}
where $\bm{\theta} \in \mathbb{R}^{L}$ is a vector of real-valued parameters $\theta^l$, $0\leq l \leq L$, where $L$ is the number of layers in the VQC. The resulting quantum state $\ket{\psi(\bm{\theta})}$ when the VQC is applied to the initial state is given by $\ket{\psi(\bm{\theta})} = U(\bm{\theta})\ket{0}$. A learned quantum function $f: \mathbb{C} \mapsto \mathbb{R}$ can then be given as
\begin{align}
    f(\ket{\psi(\bm{\theta})}, \hat{\bm{O}}) &= \bra{\psi(\bm{\theta})} \hat{\bm{O}} \ket{\psi(\bm{\theta})}\\
    \quad &= \bra{0}U^{\dagger}(\bm{\theta}) \hat{\bm{O}} U(\bm{\theta})\ket{0},
    \label{eq:cost_fn}
\end{align}
where $\hat{\bm{O}}$ is Hermitian matrix called the \emph{quantum observable}. The function $f$ can be thought of as computing the expectation value of the resulting quantum state in the eigenbasis of the observable. A VQA tries to minimize a cost function 
\[
C(\bm{\theta}) = \argmin_{\bm{\theta}} f(\ket{\psi(\bm{\theta})}, \hat{\bm{O}}).
\]
\fi

\subsection{Adaptive Threshold Detection}




 \ifisjournal
 We now introduce the component-wise adaptive threshold detection that forms one of the key features of our algorithm. Before we elaborate on the specific update rule, we first take a deeper look at the optimization process. For a coordinate wise optimizer (i.e. an optimizer which updates each components independently)
 , we can realize the following update rule from iteration $t$ to $t+1$:
 \begin{equation}
     \theta^{l}_{t+1} = \theta^{l}_t - \eta \cdot \frac{\partial C(\bm{\theta})}{\partial \theta^l_t},
     \label{eq:coordinate_update}
 \end{equation}
 where $\theta^{l}_t$ is the $l^{th}$ component of the parameter vector  $\bm{\theta}_t$ (that denotes $\bm{\theta}$ at iteration $t$), $\eta$ is the learning rate and $\frac{\partial C(\bm{\theta})}{\partial \theta^l_t}$ is the $l^{th}$ component of the gradient vector at iteration $t$. Each parameter component has different outputs from the gradient component and hence for components $\theta^{l}$ and $\theta^{l+k}$ for some layer $l, l+k$, the optimization trajectory $\{\theta^{l}_{0}, \theta^{l}_1 \dots \theta^{l}_T \}$ is different from optimization trajectory of $\{\theta^{l+k}_0, \theta^{l+k}_1 \dots \theta^{l+k}_T \}$. This difference leads us to formulate the \emph{independent parameter hypothesis}\footnote{Here ``independence" does not imply conditional independence.} that motivates this work: Each parameter in a VQC follows an independent trajectory and hence requires a threshold that is specific to the trajectory of that component. 

In order to adaptively determine the threshold for each component we initialize $\bm{\tau} \in \mathbb{R}^{n}$ for $n$ parameters with equal weights on all components. During the course of the optimization, we access the gradients $\nabla C(\bm{\theta})$ and update the threshold as
\begin{equation}
    \bm{\tau'} = \bm{\tau} - \bm{\tau} \odot |\nabla C(\bm{\theta})|
    \label{eq:tau_update}
\end{equation}

Where $|\nabla C(\bm{\theta})|$ is the absolute magnitude of the gradients. The absolute value of the gradients encodes the importance of the particular component towards the overall cost function and the update rule in Equation~\ref{eq:tau_update} yields a threshold that will prune the least important components since that's where the majority of the threshold reduction will occur. 

\fi


\subsection{Better Pruning Criteria}

\ifisjournal
In the proposed algorithm we perform the actual pruning (i.e., building a list of indices that indicate the components having the least impact towards the objective function) at fixed intervals of the optimization procedure. The length of this interval is governed by a user supplied variable $w$.

We have earlier mentioned that most pruning algorithms use the magnitude of gradient as a decision variable for pruning parameter components i.e. if $|\partial^j_\theta C(\bm{\theta}_t)|$ denotes the magnitude of $j^{th}$ component of the gradient and $\gamma$ is a scalar threshold, then the $j^{th}$ component is pruned if
\[
|\partial^j_\theta C(\bm{\theta}_t)| < \gamma.
\]
In contrast, we choose to profile a different quantity - the absolute difference in magnitude of gradients at two successive time steps $|\nabla C(\bm{\theta}_{t+1}) - \nabla C(\bm{\theta}_t)|$. We accumulate this quantity over a window size $w$ and make the decision to prune the $j^{th}$ component if the following equation holds:

\[
\sum_{t=(\kappa-1)w}^{\kappa w}|\partial^j_\theta C(\bm{\theta}_{t+1}) - \partial^j_\theta C(\bm{\theta}_t)| < \tau^j, \kappa=1, 2, \dots
\]
where we have defined $\partial^j_\theta C(\bm{\theta}) = \frac{\partial C(\bm{\theta})}{\partial \theta^j}$ and $\kappa$ is the index of current accumulation step. It can be shown that 
\[
|\nabla C(\bm{\theta}_{t+1}) - \nabla C(\bm{\theta}_t)| \approx \nabla^2 C(\bm{\theta}).
\]
Here $\nabla^2 C(\bm{\theta})$ is the Hessian of the cost function and it has been shown~\cite{lecun1989optimal, hassibi1993optimal} to be a much more useful quantity in pruning parameters. \emph{The strength of our method is that we approximate this Hessian without explicitly computing it during the optimization procedure and thus automatically provide a more nuanced criteria for pruning redundant parameters.}
\fi

\subsection{Overall Algorithm}
The overall flow of the QAdaPrune algorithm is shown in Algorithm~\ref{alg:qadaprune_alg}. Our algorithm is designed to be integrated with the normal VQA optimization procedure with minimal overhead. 

The training process proceeds with the instantiation of the threshold vector $\tau \in \{\frac{1}{n}\}^n$ for $n$ parameters. Over a window size $w$, the buffer accumulates the successive difference of the gradients. Once full (i.e. when timestep is an integral multiple of $w$), we build a saliency list of indices that are less than their corresponding thresholds. The parameters corresponding to those indices are temporarily frozen (i.e. they do not receive gradient updates) and the buffer is reset to current time stamp's gradient (Line 10). By default, our algorithm tries to freeze all indices in the saliency list. However, we realize that this mode may be too ``aggressive" for certain tasks or datasets. Hence, we also include an input parameter $k$ that allows the user to adjust how many parameters to prune at a time. If the number of indices are less than $k$ we apply the default pruning, otherwise we randomly select $k$ indices from the saliency list and freeze them in their current state. At the end of the training process, we are left with at least $k$ frozen parameters, which can later be transpiled into frozen rotation gates.

\section{Computational Experiments}\label{sec:expts}

In this section, we will demonstrate the efficacy of our approach using two distinct classification datasets. Broadly speaking, our technique successfully identifies and eliminates redundant  parameters while maintaining a negligible effect on the QNN's validation accuracy. 
The source code and results are available at \url{https://github.com/aicaffeinelife/QAdaPrune.git}.
\begin{figure}
    \centering
    \includegraphics[width=\columnwidth]{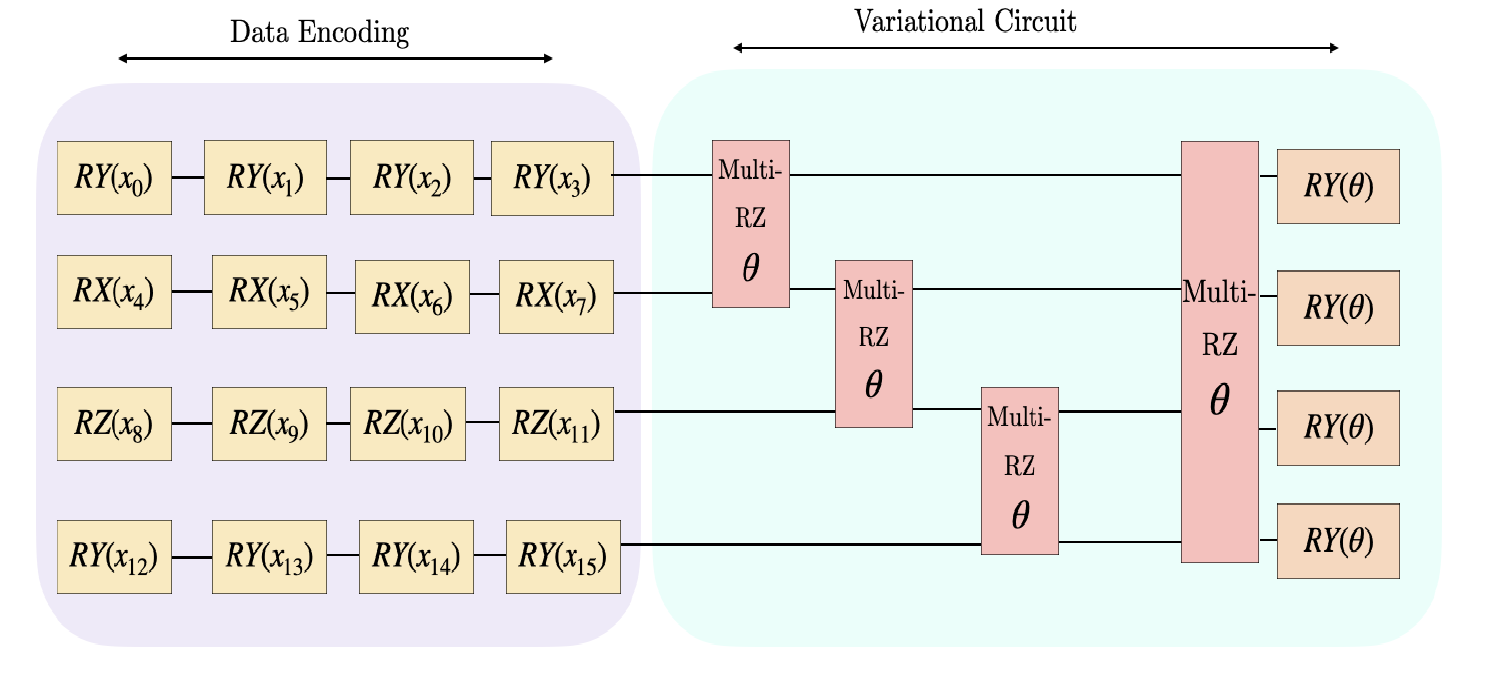}
    \caption{A diagram of the QNN used in our experiments. The circuit is viewed left to right with data encoding being applied to the input before the variational circuit. A measurement is applied at the end of the circuit to produce the predictions. (Best viewed in color).}
    \label{fig:qnn_ansatz}
\end{figure}

We perform experiments\footnote{We note that these are the realistic size experiments that can be found in the literature and be conducted on quantum simulators. The real quantum machines are both slower and more expensive to use.} on the MNIST~\cite{lecun2012mnist} and FashionMNIST~\cite{xiao2017/online} datasets. For each of the datasets we create a subset where we take 500 front images from the training set and 300 randomly chosen images as the validation set. All images are first center-cropped to $24 \times 24$ and then down sampled to $4 \times 4$ image sizes. We consider a binary classification task for both datasets. In the case of MNIST, the QNN learns to discriminate between 3 and 6 digits and in the case of FashionMNIST, the task is to learn the difference between a dress and a shirt. Our QNN circuit consists of a data encoding sub-circuit and variational sub-circuit. 

The circuits are shown in Figure~\ref{fig:qnn_ansatz}. The data encoding sub-circuit is responsible for creating the input quantum state from the given image. To accomplish this we  embed each pixel row column in a single qubit by applying a sequence of RY, RX, RZ, RY gates. The resulting quantum state $\ket{\psi}_{img}$ is then passed to the information processing sub-circuit which consists of alternating RZZ gates followed by a layer of RY rotation. At the end of the circuit we perform a measurement on all four qubits with the Pauli-Z observable. This measurement collapses the output state from each qubit $\ket{\psi}_{out}$ to be in $\{-1, +1\}$. For the binary classification problem, we follow the method in~\cite{wang2022qoc} and reduce the four-valued output vector to a two valued one by summing the first two and the last two components. We use the binary cross entropy loss for training the quantum circuits:

\begin{equation}\label{eq:bce_loss}
C(\bm{\theta}) = \sum_{i=0}^M y_i\log(\hat{y}_i) + (1 - y_i)  \log(1 - \hat{y}_i),    
\end{equation}
where $\hat{y}_i$ is the output vector produced by the QNN and $M$ is the size of the dataset.

\subsection{Main Results}

\subsubsection{Classification}
\begin{table}[]
    \centering
    
    \begin{tabular}{c |c |c}
    \toprule
        Method &  MNIST-2 & FashionMNIST-2 \\
         \midrule
        Q.C. - noiseless  & 90.33 & 87.67 \\ 
        Q.C. - QAdaPrune & 90.00 & 87.00 \\  
        \hline 
        Q.C.- limited shots & 86.33 & 87.60 \\
        Q.C.- QAdaPrune-limited shots& 88.67 & 87.30 \\

    \bottomrule
    \end{tabular}
    \caption{Validation accuracy on MNIST and FashionMNIST datasets for two class classification datasets. The pruned version of the Q.C. has at least 25\% pruned parameters.}
    \label{tab:results_best_qnn}
\end{table}

Table~\ref{tab:results_best_qnn} shows the main results for classification task on the two datasets. We compare our method in four different settings: (1) QNN with no pruning and analytical gradient computation, (2) QNN with QAdaPrune procedure assuming analytical gradients, (3) QNN trained with no pruning and noisy gradient computation and (4) QNN with pruning applied assuming noisy gradients. It is necessary to separate noisy and noiseless gradients cases because in the former case the gradients are estimated assuming there are infinite ``shots" (i.e., number of qubits available for computation) while in the more realistic latter case the number of shots are provided by the quantum device leading to noisy estimation of gradients. All experiments were performed using the Pennylane~\cite{bergholm2018pennylane} software using the quantum device simulators provided by the library. In the limited shots setting, we limit the number of shots to $10000$.

\begin{figure*}[t]
    \centering
    \includegraphics[width=2\columnwidth]{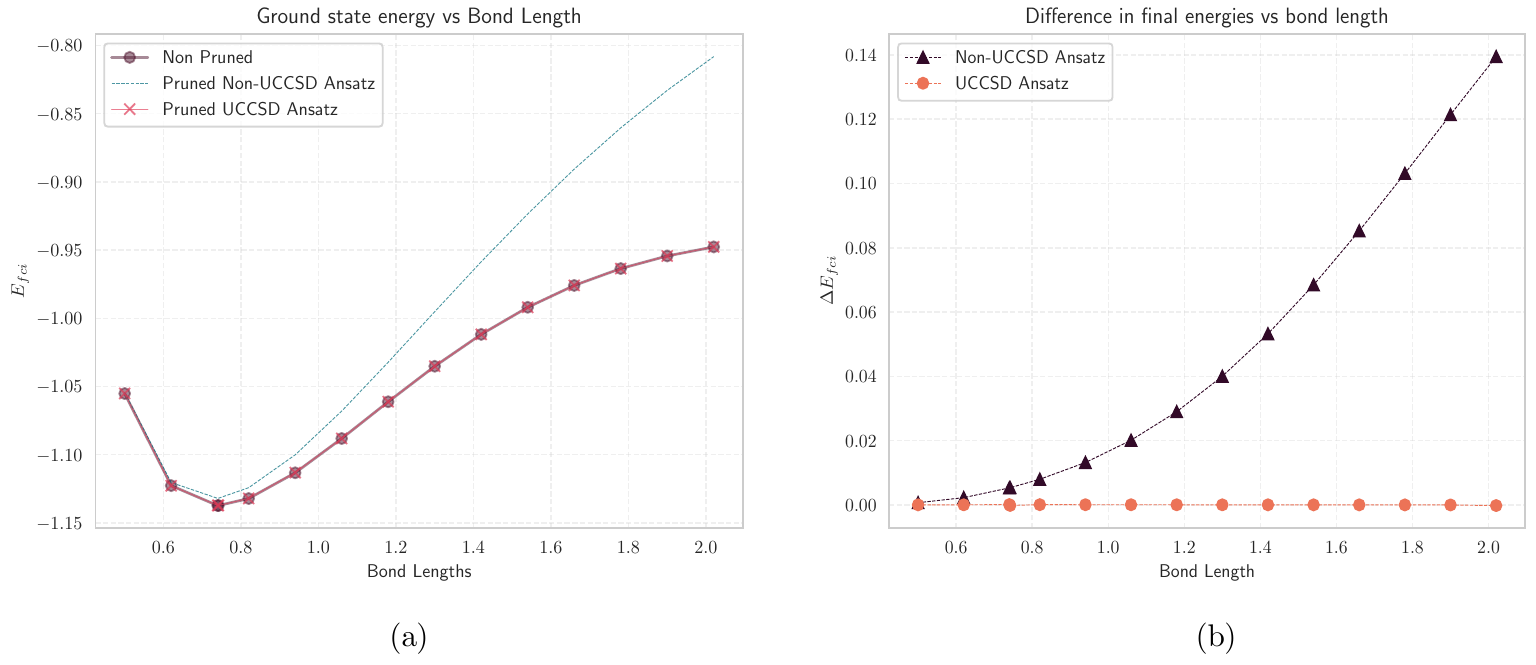}
    \caption{Performance of VQE with UCCSD and customized ansatz with pruning and non-pruning setting. Left: Ground state energies obtained by different ansatz. Right: Difference in the ground state energy between pruned and unpruned cases.}
    \label{fig:vqe_energy_comparison}
\end{figure*}


In instances with analytical (noiseless) gradients for both tasks, the QAdaPrune method achieves comparable accuracy to the unpruned scenario while using at least 25\% fewer parameters. This suggests that QNNs, much like their classical counterparts, often have redundant input representations across various pathways from input to output. When gradients are estimated with limited shots, the QNN's accuracy experiences only a slight dip in performance upon pruning. This minor decline in accuracy in both the noiseless and noisy scenarios can also be viewed as a regularization effect, preventing the QNN from overfitting to dataset noise. Given that QNNs are highly sensitive to any form of noise, whether in the dataset or during qubit preparation, our pruning method offers an added advantage of stabilizing the QNN training process.

\subsubsection{Finding ground state energy}

\begin{figure}
    \centering
    \includegraphics[width=\columnwidth]{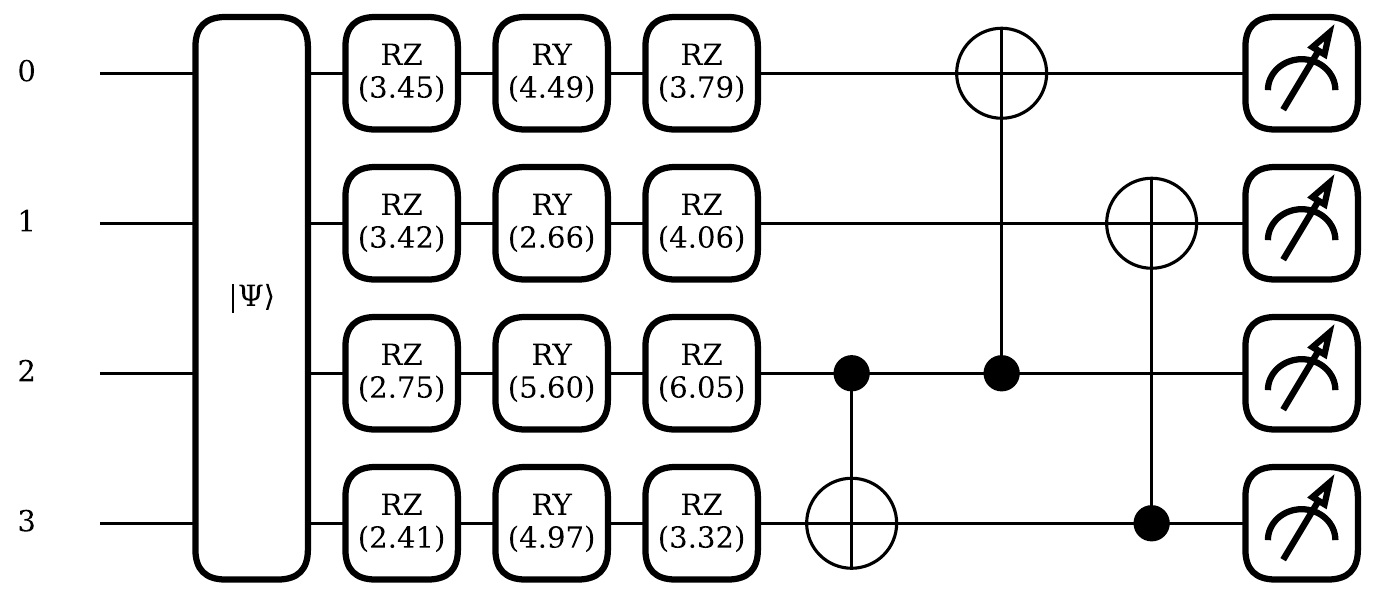}
    \caption{Non-UCCSD Ansatz initialized with random parameters used in our VQE experiments.}
    \label{fig:vqe_circuit}
\end{figure}

To further investigate our pruning strategy, we implement the problem of estimating the ground state energy of molecular Hamiltonian with the use of Variational Quantum Eignesovler~\cite{VQE}. In this experiment, the pruning scheme is implemented with the use of UCCSD ansatz and a customized ansatz to estimate the groundstate energy of $H2$ (hydrogen) Hamiltonian. The custom ansatz with an initial parameterization is shown in Figure~\ref{fig:vqe_circuit}. For both ansatz, the initial state is set to HF state of $H2$ (hydrogen) Hamiltonian which is [1, 1, 0, 0]. For the optimization, the Adam optimizer \cite{kingma2014adam} is used with the learning rate $1e^{-3}$ and $500$ max iterations.  We used the a window size of $5$ for both UCCSD and non-UCCSD ansatz.  

\begin{figure}
    \centering
    \includegraphics[width=\columnwidth,height=5cm]{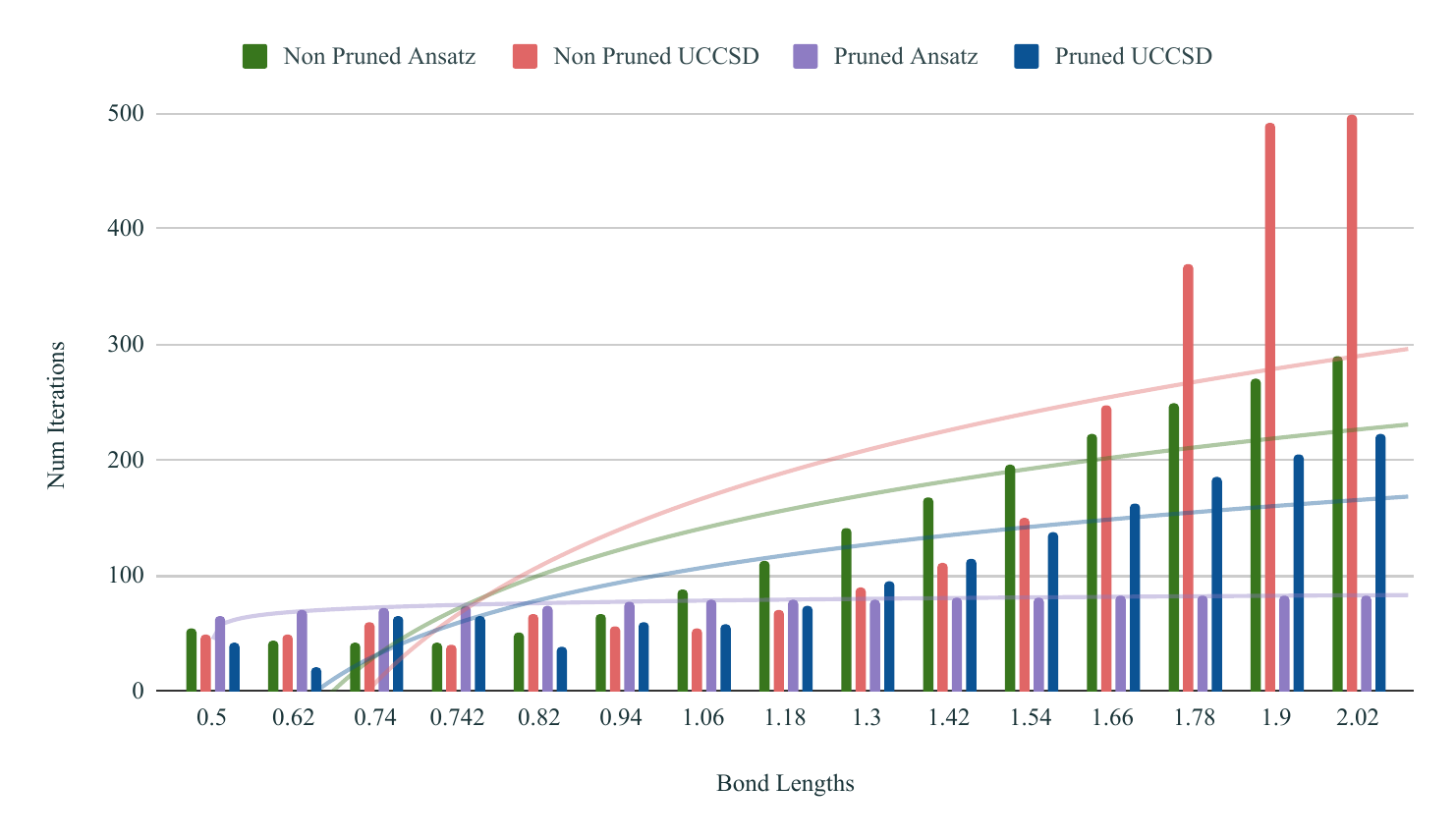}
    \caption{The number of iterations taken to converge for the VQE optimization process for non pruned and pruned cases}
    \label{fig:ite_cases}
\end{figure}

Figure \ref{fig:vqe_energy_comparison}(a) demonstrates final ground state energies obtained by the non pruned and pruned ansatz for various bond lengths. Figure~\ref{fig:vqe_energy_comparison}(b) shows the difference in the ground state energies between the pruned and non-pruned cases. We can immediately see that there is no performance degradation between pruned and non-pruned UCCSD ansatz. However, the non-UCCSD ansatz suffers a significant decrease in performance after pruning. However, for small bond lengths ($\leq 1.2 \AA$), there is no significant decrease in the pruned and non-pruned energies. This shows that our algorithm with a static window size can be easily applied to molecules up to a certain limit without suffering a decrease in performance. We further profiled the number of iterations it took for the pruned and non-pruned ansatze for both cases. The results of the experiment are shown in Figure~\ref{fig:ite_cases}. We can see that even though our pruning algorithm has an overhead of computing the saliency, the number of iterations taken to converge for each bond length is significantly less than the non-pruned case. This indicates that pruning has a beneficial effect of introducing constructive noise during the training process. This constructive noise has been shown to help stochastic optimization processes converge faster~\cite{ge2015escaping}.

\ifisjournal
\subsection{Ablation Results}

In our algorithm the main hyperparameter is the gradient accumulation window size $w$. We explore the impact of this hyperparameter by varying the window size from $1$ to $10$. When the accumulation size is $1$, our algorithm becomes an online pruning procedure that builds pruning indices at each training step. In our experiments we noted that this online pruning mode can have a beneficial regularization effect on the QNN. Figure~\ref{fig:ablation_results} shows the  best validation accuracy with varying window sizes for ideal and limited shot settings described earlier.

\begin{figure}[t]
    \centering
    \includegraphics[width=\columnwidth,height=3.5cm]{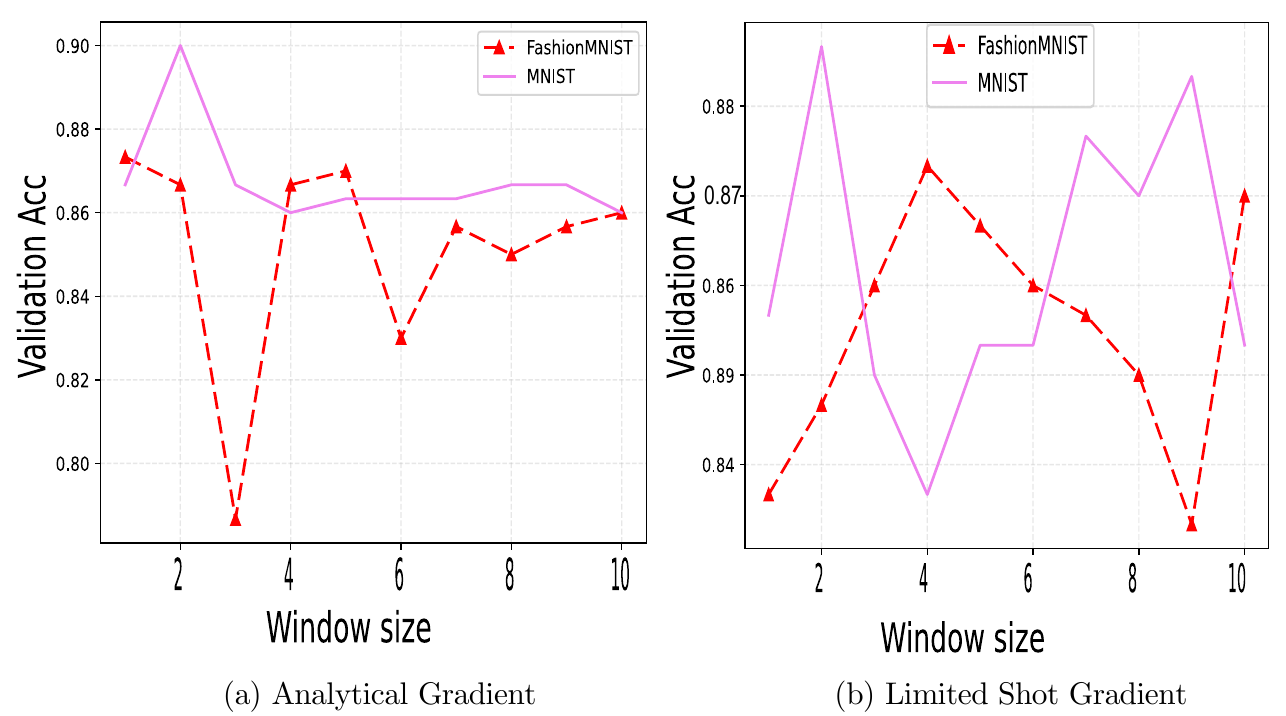}
    \caption{Ablation results with different window sizes for analytical and limited shot settings.}
    \label{fig:ablation_results}
\end{figure}

\begin{figure}[h]
    \centering
    \includegraphics[width=\columnwidth, height=3.5cm]{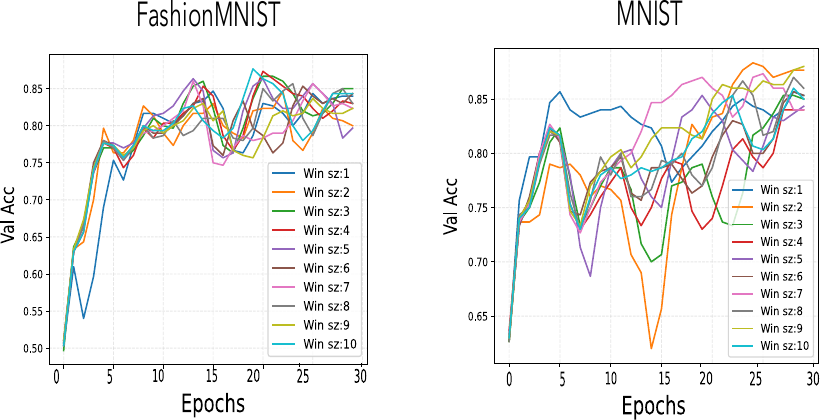}
    \caption{Validation accuracy over training epochs with varying window sizes}
    \label{fig:ablation_acc_hist_win_sz}
\end{figure}

In the case where we have access to noiseless (i.e. analytical) gradients the optimal window size for MNIST task is $2$ and $4$ for FashionMNIST. The case of noisy (estimated from limited trials) has much more variance and we see that accuracy can significantly vary with window size. Even so, we find that the same window sizes that worked for the two tasks in the noiseless case are also the best performing ones in the noisy case as well. Figure~\ref{fig:ablation_acc_hist_win_sz} shows the validation accuracy over the epochs for different window sizes for both tasks when gradients are estimated in a noisy fashion. We can see that for FashionMNIST dataset, the noise does not significantly affect the performance of the QNN. On the other hand, for the MNIST dataset the QNN is highly impacted by noise. However, since the QNN ultimately converges to similar accuracy for every window size, we can conclude that pruning can have beneficial regularization effects similar to dropout~\cite{srivastava2014dropout} in classical deep neural networks. 
\fi
\section{Related Work}

\begin{table}[h]
    \centering
    \begin{tabular}{p{0.11\textwidth}|p{.05\textwidth}|p{.09\textwidth}|p{.09\textwidth}}
    \toprule
    Method & Adaptive & Threshold & Pruning Criteria \\
    \midrule
    Quantum-on-Chip~\cite{wang2022qoc}& \ding{55} & Hyperparameter-defined & Gradient Magnitude \\  
    \hline
    QNN Compression~\cite{hu2022quantum} & \ding{55} & Varies with quant. angle & Gate property defined \\ 
    \hline

    Adaptive Pruning~\cite{sim2021adaptive} & \ding{51} & User-defined & Gradient Magnitude \\ 
    \hline
    Ours & \ding{51} & Adaptive and learned & Approximated Hessian
    \end{tabular}
    \caption{Algorithmic differences between different pruning methods in quantum computing.}
    \label{tab:relwrk}
\end{table}

Table~\ref{tab:relwrk} summarizes the key differences between pruning methods that have been proposed for quantum circuits in literature and our algorithm. Most existing pruning algorithms either consider a static threshold that is input as a hyperparameter or rely on a hyperparameter to serve as a proxy threshold. For instance, Hu~\emph{et al.}~\cite{hu2022quantum} propose using integral multiples of $\pi$ for rotation angles as thresholds and prune them if the evaluate to identity at those angles. Wang~\emph{et al.} use a hyperparameter $r$ to randomly select the number of parameters to prune over a pruning window $w_p$. In contrast, our method adjusts the threshold individually for each component based on the amount of contribution they make to the cost function.  Another aspect of existing pruning algorithms is that they require a large number of tunable hyperparameters to effectively prune a given QNN. For instance, Sim~\emph{et al.}~\cite{sim2021adaptive} propose a pruning method with four hyperparameter and rely on heuristics to prune and regrow parameters. In contrast we require at most two parameters for our algorithm which can be easily estimated before training the QNN. The final key difference of our method from all previous methods is that we are the first ones to consider approximating a Hessian of the gradient in a dynamic manner and prune parameters based on that information.

\section{Conclusion}\label{sec:conclusion}
In this paper we have introduced QAdaPrune - an adaptive algorithm that is able to dynamically determine per-parameter component thresholds and propose novel criteria to determine if the parameters need to be pruned. This step is extremely important to eliminate current obstacles in deployment of quantum machine learning and optimization algorithms. Our experiments on different tasks show that our algorithm is a general purpose and can be employed with minimal changes to the overall optimization procedure. More importantly, besides using at most two hyperparameters we do not make any assumptions about the pruning process. We also profiled the run time on the tasks considered in this paper. On FashionMNIST task a normal epoch (without pruning) takes about 1 minute on average. With pruning, a training epoch takes 1 minute and 41 seconds. On MNIST, both normal and pruning procedures take about 1 minute on average. Thus, our procedure adds minimal overhead to the training process.

In this paper we have not considered the impact of system noise on the performance of QAdaPrune. We believe that a future research direction that investigates this question will yield interesting insights about the effects of pruning on training of a QNN with noisy quantum hardware. The choice of the initializing distribution (e.g., beta distribution \cite{kulshrestha2022beinit}) is another direction to study in combination with smart pruning to verify its impact on  barren plateaus. Another viable future research direction to explore is to introduce QAdaPrune as a preprocessing step to identify redundant parameters before transpiling a deep VQC into a more efficient circuit.  


\section{Acknowledgements}
This work was supported  in part with funding from the Defense Advanced Research Projects Agency (DARPA), under the ONISQ program. The views, opinions and/or  findings expressed are those of the author and should not be interpreted as representing the official views or policies  of the Department of Defense or the U.S.Government.

This research was supported in part through the use of DARWIN computing system: DARWIN - A Resource for Computational and Data-intensive Research at the University of Delaware and in the Delaware Region, which is supported by NSF Grant \#1919839.


\bibliographystyle{IEEEtran}
\bibliography{ref}
\end{document}